\documentclass[lettersize,journal]{IEEEtran}
\usepackage{amsmath,amsfonts,amssymb}

\DeclareMathOperator{\prob}{Pr}
\newcommand\mperiod[1][\rlap]{#1{\;.}}
\newcommand\mcomma[1][\rlap]{#1{\;}} 
\newcommand\msemi[1][\rlap]{#1{\;;}}

\DeclareMathOperator*{\cdf}{cdf}

\DeclareMathOperator*{\ccdf}{ccdf}
\DeclareMathOperator{\Agree}{Agree}

\usepackage{algorithmic}
\usepackage{algorithm}
\usepackage{array}
\usepackage[caption=false,font=normalsize,labelfont=sf,textfont=sf]{subfig}
\usepackage{textcomp}
\usepackage{stfloats}
\usepackage{url}
\usepackage{verbatim}
\usepackage{graphicx}
\usepackage{fontawesome5}
\usepackage[capitalise]{cleveref}
\newtheorem{theorem}{Theorem}
\crefname{theorem}{Theorem}{Theorems}

\usepackage{tikz}
\usetikzlibrary{shapes.geometric}

\usepackage[backend=biber,style=ieee,sorting=none]{biblatex}
\begin{document}

\title{Spoofing Missed-Detection Bounds for PRF GNSS Ranging Authentication Under AWGN Models}

\author{Jason Anderson,~\IEEEmembership{Xona Space Systems}
\thanks{This is a preprint.}}

\markboth{Preprint}%
{Anderson: Spoofing Missed-Detection Bounds for PRF GNSS Ranging Authentication Under AWGN Models}


\maketitle

\begin{abstract}
Pseudorandom-function (PRF) ranging codes, such as those used in Galileo's encrypted E6-C under the Signal Authentication Service (SAS), enable a receiver to authenticate pseudoranges once the PRF secret is revealed.
This work bounds how much authentication security the receiver obtains under Additive White Gaussian Noise (AWGN) assumptions.
Against a spoofer that does not estimate the code before submitting its forgery, PRF security makes the forged correlation zero-mean up to the security of the underlying PRF, allowing integration time and C/N$_0$ to mostly determine probability of missed detection (PMD) and probability of false alarm (PFA).
Against such a spoofer at a conservative 30 dB-Hz, 400 ms of E6-C aggregation certifies a PMD below $2^{-128}$ (plus any PRF advantage).
For a spoofer that estimates chips before submitting a forgery, I derive the receiving-antenna gain at which authentication security breaks, which is about 12 dB for E6-C for the adversaries modeled.
This work can be used to design a PRF GNSS ranging code protocol and a receiver capable of correctly asserting PRF ranging security assuming an AWGN model.
\end{abstract}


\section{Introduction}
    \IEEEPARstart{C}{ryptographic} authentication for satellite navigation ranging is progressing quickly from concept to real-world deployment.
    Galileo's Signal Authentication Service (SAS) has started orbit testing with encrypted E6-C signals~\cite{ieeeSAS,testingSAS}.
    And, Xona Space Systems broadcast its first watermarked signal in July 2025~\cite{worldsfirstwatermarkrange}.
    This work comes just in time as Global Navigation Satellite System (GNSS) signal spoofing becomes a common nuisance in conflict zones and nearby public infrastructure.

    GNSS signals typically provide two signal component types: data and ranging.
    Data components typically provide information related to broadcasting satellites' position and clock bias.
    Ranging components enable receivers to measure their range to satellites by transmitting a high-auto-correlative, low-cross-correlative ranging code.
    These two components together enable receivers to solve a trilateration problem to deduce their position and time.
    Cryptographic ranging authentication involves augmenting the ranging components with cryptographic elements~\cite{Scott2003}.
    The cryptographic construction asserts that these elements are not predictable by GNSS spoofers before broadcast.
    When a receiver observes these cryptographic elements, it establishes trust in the measured pseudoranges that determine the receiver's position and time.

    Cryptographic ranging authentication comes in two forms.
    The first involves inducing a watermark within the ranging code~\cite{worldsfirstwatermarkrange,Anderson2017}.
    The watermark enacts a cryptographic perturbation on the ranging code chips\footnote{Each bit in a GNSS ranging code is called a chip rather than a bit.} that is designed to be small.
    And because the watermark is a small perturbation, receivers can still track the signal and ignore the watermark's existence.
    The second form of cryptographic ranging authentication requires the entire ranging code derive from a secure cryptographic Pseudorandom Function (PRF).
    When the ranging code derives from a secure PRF, receivers cannot track the signal until they have the cryptographic seeds used to generate the PRF ranging code.

    Both watermark and PRF ranging authentication systems enact a bit-commitment authentication protocol, usually under a Timed Efficient Stream Loss-tolerant Authentication (TESLA) framework~\cite{perrig2005timed}.
    Xona's Pulsar, GPS's Chimera, and Galileo's SAS utilize TESLA~\cite{worldsfirstwatermarkrange, chimeraicdTESLA,ieeeSAS}.
    The authentication procedure can be briefly summarized as follows.
    First the GNSS provider commits to a secret from which the watermarked or the PRF ranging code derives.
    Second, after a delay, the GNSS provider will reveal that commitment secret to receiver.
    Third, the receiver can retrieve the corresponding radio measurements in the past and process the watermark or PRF ranging code to determine a pseudorange's authenticity utilizing the revealed secret.

    For a cryptographic ranging authentication system to be useful, there must be a receiver signal processing procedure that is sensitive and specific.
    The probability a spoofer could produce a signal that would fool a receiver's authenticity determination, called the Probability of Missed Detection (PMD), must be small.
    Moreover, the probability an authentic signal would raise an alarm, called the Probability of False Alarm (PFA), must also be small.
    For the watermark case, when the watermark is a Combinatorial Watermark, there exists a receiver processing procedure that can compute the PMD and PFA and then assert cryptographic authentication security under Additive Gaussian White Noise (AWGN) models~\cite{anderson2024gnsscrypto}.
    With the mathematical PMD and PFA models, the watermark degradation can be minimized while achieving PMD and PFA levels similar to standard security guidelines.

    For the PRF case, several works propose receiver processing statistics based on the pseudorange differences between authenticating PRF pseudoranges and non-authenticated pseudoranges to authenticate the non-authenticated pseudoranges~\cite{testingSAS,sasReceiver}.
    However, environmental effects unrelated to spoofing threats interfere with this methodology's specificity (i.e., the observed PFA), nor do they provide mathematical proof of security by bounding the PMD.
    Previous work utilized statistic arguments on multiple chip measurements~\cite{Psiaki2016,scerpower}.
    However, they do not exploit the relationship among many unpredictable chips and the quickly-decaying tails of the binomial distribution to provide mathematical arguments of the levels of standard cryptographic security.
    This work provides the equivalent to that provided for Combinatorial Watermark ranging by \cite{anderson2024gnsscrypto,Andersonnavi.655,Andersonnavi.696} for the PRF ranging code case.
    After asserting standard cryptographic security levels utilizing this work, receiver thresholds on pseudorange differences can be expanded to increase the processing specificity without compromising security.

    This work provides a signal processing procedure and demonstrates how to bound the PMD and PFA of receiver statistics of PRF ranges to standard cryptographic security levels (e.g., 128-bit security).
    In this work, a security level of $s$ bits means the PMD is bounded by $2^{-s}$, so that an adversary must submit around $2^s$ forgeries in expectation to succeed once.
    Unlike the conventional interpretation of $2^s$ offline operations, each forgery attempt against PRF ranging requires a physical transmission that the victim receiver and local authorities may observe.

    I apply multiple security models to these security claims and provide simulated results of adversary capabilities.
    The principal mathematical results are each verified with Monte Carlo experiments.
    And each of the results are applied to Galileo's SAS with Galileo's E6-C service, an upcoming PRF ranging authentication provider.

    \subsection{PRF Ranging}

        GNSS ranging codes are designed to have high autocorrelation and low cross-correlation.
        The receiver will use a replica of the ranging code to shift-correlate with its observed radio signal.
        The high autocorrelation enables the receivers to lift the signal from the noise, and the sharp autocorrelation peak enables the receiver to measure a precise signal arrival time forming a pseudorange.
        At least four satellite pseudoranges enable a receiver to deduce its position and time via trilateration.

        Random codes can serve as ranging codes since random sequences have high autocorrelation and low cross-correlation.
        Moreover, pseudorandom code sequences, sequences that appear random but actually derive from a deterministic process, also suffice.
        When a ranging code derives from a cryptographic PRF, the ranging code inherits cryptographic security properties.
        For instance, when a ranging code derives from a cryptographic PRF utilizing a secret seed selected uniformly randomly, then the ranging code admits no efficient algorithm that can distinguish it from a truly random sequence nor predict future chips with probability better than random guessing.
        Before a PRF ranging code's broadcast, their security limits adversarial attempts to forge ranging signals to that of random guesses.
        PRF ranging codes can be derived multiple ways, including via encryption or cryptographic hashing.
        PRF ranging exists in practice today with encrypted GNSS signal's such as GPS's M-Code and Galileo's PRS.

        Galileo's SAS will enact PRF ranging authentication with Galileo's encrypted E6-C signal with assistance from Galileo's Open Service Navigation Message Authentication (OSNMA) on Galileo's E1-B signal~\cite{fernandez2024galileoSAS,TerrisGallego2024}.
        The design can be summarized as follows.
        Galileo will encrypt the ranging codes of its E6-C signal and never disclose the deriving key.
        Galileo will perform a second encryption on the E6-C ranging codes utilizing the undisclosed TESLA keys from OSNMA.
        These second-encrypted ranging codes will be available for download from a server to any receiver.
        As OSNMA reveals TESLA keys, the receiver will be able decrypt the second-encrypted E6-C ranging codes and perform post processing of the E6-C signal to determine authenticity.
        Receivers can utilize the derivations and procedures of this work to determine authenticity with high confidence.

    \subsection{Chip Estimation Theory} \label{sec:chip-est-theory}

        Ranging signals can be modeled as binary keyed code, usually Binary Phase-Shift Keyed (BPSK).
        Each broadcast bit, mapped to $-1$ or $1$, is called a chip.
        When the ranging code derives from a secure PRF, the adversary cannot predict the chips before their broadcast.
        But some adversaries considered in \cref{sec:adv-models} can measure the chips after broadcast.
        Typically, GNSS ranging signals are below the thermal noise floor, which makes measuring individual chips difficult without large directional antenna.

        Given the noise power $\sigma^2$ and the signal power $P$, a single  measurement of the BPSK baseband (i.e., after carrier removal) signal $S$ is
        \begin{align}
            N &\sim \mathcal{N}(0, \sigma^2) \msemi \\
            S &= \pm \sqrt P + N \mperiod \label{eq:single-chip}
        \end{align}
        With this model, the probability of successfully measuring a chip correctly $p$ is
        \begin{align}
            p &= \cdf_N\left(\sqrt P\right) \mperiod \label{eq:chip-est-correct}
        \end{align}
        This chip estimation model is simple but effective, serving the purposes of this work.

    \subsection{GNSS Signal and Receiver Processing Models}\label{sec:radio-model}

        GNSS signals typically involve a baseband signal modulated with a carrier wave.
        They are also typically subdivided into a discrete ranging code units, for which we let $n$ be the number of chips per unit and $T$ be time length per unit.
        For instance, Galileo E6-C utilizes $n=5115$ chip ranging codes that repeat every $T=1$ ms.
        Each millisecond ranging code within Galileo E6-C can become a PRF ranging code at will when encryption is turned on.

        The mathematics of this work begin after the receiver has removed the carrier wave with a tracking loop.
        Later in this section, I discuss how I account for the effect of tracking errors.
        Therefore, let the GNSS authentic baseband signal be
        \begin{align}
            S^\text{auth} &= \sqrt P R^\text{PRF} + N \mperiod \label{eq:auth-signal}
        \end{align}
        In \cref{eq:auth-signal}, $S$, $R$, and $N$ are each vectors that span a single ranging code.
        When the receiver sampling frequency is $F$, then the length of each is $FT$.
        $R^\text{PRF} \in \{-1, 1\}^{FT}$ is the resampled {\em replica} of the PRF sequence.
        $P$ is the power of the signal, which for now is constant per ranging code, but this assumption is relaxed in \cref{sec:pscer-adv}.

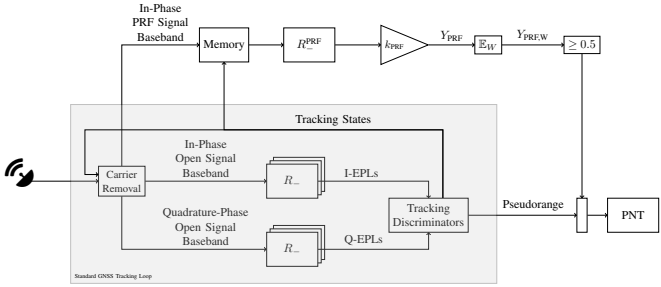
\begin{figure}[!t]
            \centering
            \resizebox{\linewidth}{!}{
            \begin{tikzpicture}
                \node (antIcon) at (-1,0.25) {\Huge \rotatebox{90}{\faSatelliteDish}};
                \node (ant) at (-1,0) {};
                \node[draw, rectangle, align=center] (osc) at (2,0) {\small Carrier\\\small Removal};
                \draw[->] (ant) -- (osc);

                \node[draw, rectangle, minimum height=1cm, minimum width=1.5cm, fill=white] at (7.2, 0.2) {};
                \node[draw, rectangle, minimum height=1cm, minimum width=1.5cm, fill=white] at (7.1, 0.1) {};
                \node[draw, rectangle, minimum height=1cm, minimum width=1.5cm, fill=white] (matchI) at (7, 0) {$R_-$};
                \node[draw, rectangle, minimum height=1cm, minimum width=1.5cm, fill=white] at (7.2,-1.8) {};
                \node[draw, rectangle, minimum height=1cm, minimum width=1.5cm, fill=white] at (7.1,-1.9) {};
                \node[draw, rectangle, minimum height=1cm, minimum width=1.5cm, fill=white] (matchQ) at (7,-2) {$R_-$};
                \draw[->] (osc) -- node[anchor=south, align=center] {In-Phase\\Open Signal\\Baseband} (matchI);
                \draw[->] (osc) |- node[pos=0.79, anchor=south, align=center] {Quadrature-Phase\\Open Signal\\Baseband} (matchQ);

                \node[draw, rectangle, minimum height=1cm, minimum width=1.5cm, fill=white, align=center] (TD) at (11, -1) {Tracking\\Discriminators};
                \draw[->] (matchI) -| node[pos=0.1, anchor=south west] (IP) {I-EPLs} (TD);
                \draw[->] (matchQ) -| node[pos=0.1, anchor=south west] (IQ) {Q-EPLs} (TD);

                \node[draw, rectangle, minimum height=1cm, minimum width=1.5cm, fill=white] (PNT) at (17, -1) {PNT};
                \node[draw, rectangle, minimum height=1cm, minimum width=0.3cm, fill=white] (block) at (15.5, -1) {};
                \draw[->] (TD) -- node[anchor=south,pos=0.6] {Pseudorange} (block);
                \draw[->] (block) -- (PNT);

                \node[draw, rectangle, minimum height=1cm, minimum width=1cm] (mem) at (5, 4) {Memory};
                \draw[->] (osc) |- node[pos=0.75, anchor=south, align=center] {In-Phase\\PRF Signal\\Baseband} (mem);

                \node (RPRF) at (7.5, 4) [draw, rectangle, minimum height=1cm, minimum width=1.5cm] {$R^\text{PRF}_-$};
                \node[draw, fill=white, isosceles triangle] (kPRF) at (10, 4) {\small $k_\text{PRF}$};
                \node[draw, fill=white] at (12.75, 4) (EW) {$\mathbb{E}_W$};
                \node[draw, fill=white] at (15.5, 4) (threshold) {$\geq 0.5$};

                \draw[->] (mem) -- (RPRF);
                \draw[->] (RPRF) -- (kPRF);
                \draw[->] (kPRF) -- node[anchor=south] {$Y_\text{PRF}$} (EW);
                \draw[->] (EW) -- node[anchor=south] {$Y_\text{PRF,W}$} (threshold);
                \draw[->] (threshold) -- (block);

                \draw[fill=black!20, opacity=0.25] (0.5, 2.25) rectangle (13, -3);
                \node[anchor=south west] at (0.5, -3) {\tiny Standard GNSS Tracking Loop};

                \draw[->] ([xshift=-0.75cm]TD.north east) -- ++(0,2) -- ++(-10.5, 0) |- ([yshift=-0.25cm]osc.north west);
                \draw[->] ([xshift=-0.75cm]TD.north east) -- ++(0,2) -| node[anchor=south, pos=0.25] {Tracking States} (mem.south);

            \end{tikzpicture}
            }
            \caption{
                An example radio block diagram that describes how PRF ranging would occur.
                The gray box diagrams a typical tracking loop where early, prompt, and late (EPL) correlations are used to form a tracking loop to determine the carrier phase, doppler, and code phase.
                An unknown PRF sequence is known to be present, so the baseband samples are stored into memory awaiting the PRF replica seed.
                After distribution of the PRF seed, the receiver will do an additional correlation with the PRF replica to determine authenticity.
                This figure depicts the case where a PRF signal is used to authenticate another signal: like if the encrypted Galileo E6-C is used to authenticate the open Galileo E1-B.
                A similar authentication system could use the PRF ranges directly for Position, Navigation, and Timing (PNT).
                }
            \label{fig:radio}
        \end{figure}

        \Cref{fig:radio} provides a block diagram of a radio performing the authentication determination.
        Within the gray box of \cref{fig:radio} exists a standard GNSS tracking loop where, after carrier removal, multiple replica correlations aid a tracking loop.
        Initially, the receiver will not have knowledge of $R^\text{PRF}$, so it must receive the authenticated tracking states from a network or by tracking another synchronized signal.
        For instance, with Galileo's SAS, the Galileo's encrypted E6-B and Galileo's open E1-B signal can initially be used for tracking before observing the encrypted E6-C after the PRF secrets are revealed.
        Because Galileo's E6-B is transmitted on the same band as E6-C (as opposed to Galileo's E1-B), utilizing Galileo's E6-B tracking states will not need corrections related to the signal carrier (assuming, reasonably, Galileo can transmit E6-B and E6-C with a fixed phase delay perfectly).
        Before the receiver gets the needed secrets to derive $R^\text{PRF}$ within the E6-C signal, the receiver must store the baseband samples in memory.

        After the receiver derives $R^\text{PRF}$, it can process those stored baseband samples to determine authentication.
        Within \cref{fig:radio}, the samples from memory are fed into a linear time-invariant (LTI) matching filter $R^\text{PRF}$, then the gain from \cref{eq:k}, then an averaging filter (averaging the result of $W$ ranging codes), and then finally a threshold gate that determines authenticity.
        \begin{align}
        k_\text{PRF} &= \frac{1}{||R^\text{PRF}||_1} \frac{1}{\sqrt P} = \frac{1}{FT} \frac{1}{\sqrt P} \label{eq:k}
        \end{align}
        The construction of filter and gain from \cref{fig:radio} was judiciously chosen to enable a concise mathematical pathway to bound the PMD and PFA, which will be apparent later.

        Implicit in the filter gain from \cref{eq:k} is that the radio has a secure way (i.e., resistant to adversarial manipulation) to determine the power of the incoming signal.
        In this work, I adopt the argument from~\cite{anderson2024gnsscrypto} by assuming that a receiver will cease to function and will not attempt to authenticate when the observed C/N$_0 < 30$ dB-Hz for the band containing the PRF ranging code.
        Typical receivers coherently integrating over $T=1$ ms will need much larger C/N$_0$, meaning that 30 dB-Hz is meant to be an unreasonably low worst case.
        So throughout this work, I assume that the receiver will cease making an authentication determination if the measured C/N$_0 < 30$ dB-Hz for the band containing the PRF ranging code.
        For the security derivations, I assume that C/N$_0 = 30$ dB-Hz for the band containing the PRF ranging code.
        Because assuming this unfavorable noise condition makes the security results worse, the receiver should expect better security in reasonable operating conditions.
        But by selecting this conservative value, I claim (without rigorous proof) that I account for the uncertainty resulting from an adversary manipulating the receiver's power estimator, simultaneously jamming the receiver, or inducing tracking errors.

        Similarly, when forming a radio model useful for making authentication security arguments, it is appropriate to form a conservative, worst-case receiver radio model.
        For this work, I assume that the radio operates at the Nyquist Frequency (in addition to the unreasonable C/N$_0$ from the previous paragraph).
        For instance, for Galileo's E6-C, $F = 10.230$ MHz.
        Again, one should expect better-reported-security for a real receiver out in the field because they will be utilizing a sampling frequency greater than the required Nyquist frequency.

        \Cref{tab:notation} provides a notation table for the reader's convenience.

        \begin{table}
            \begin{tabular}{|p{0.1\linewidth}|p{0.8\linewidth}|} \hline
                \textbf{Notation} & \textbf{Definition and Short Description} \\ \hline
                $n$ & The number of chips within a single ranging code (or ranging code segment). For instance, for Galileo's E6-C, $n=5115$. \\
                $T$ & The length of time covering a single ranging code. For instance, for Galileo's E6-C, $T=1$ ms. \\
                $W$ & The number of ranging codes (or ranging code segments) that should be aggregated when making an authentication determination. \\ 
                $R$ & A replica of the ranging code with elements $-1$ and $1$, sometimes with a superscript denoting what generated the replica and subscript denoting whether it is reversed in time for a matching convolution. \\
                $F$ & The sampling rate of the radio making the authentication determination. \\
                $S$ & The signal measured over $T$ with frequency $F$. \\
                $P$ & The power of the ranging signal immediately preceding correlation. \\
                $\sigma^2$ & The noise power immediately preceding correlation. \\ 
                $p$ & The probability of correctly estimating a chip. \\ \hline
            \end{tabular}
            \caption{Notation Table} \label{tab:notation}
        \end{table}

    \subsection{Adversarial Threat Models} \label{sec:adv-models}  

        To begin defining the adversarial models of this work, first consider a simple GNSS spoofer with only a transmitting antenna for an open (i.e., not encrypted) signal.
        This adversary has access to the signal specification, and the ranging code is simply a repeated, known ranging code.
        So the adversary can utilize the specification to submit forged signals to the receiver.
        To manipulate the victim's deduced position and time, the adversary manipulates the arrival times of the forged signals.
        Now suppose the spoofed signal derives from a PRF derived from a secret unknown to anyone except the GNSS provider.
        Without the needed cryptographic secrets, the adversary cannot predict the ranging codes before the GNSS provider broadcasts them, even with possession of the signal specification.
        However, a GNSS spoofer with listening and transmitting antenna could observe and then replay the PRF sequence to the victim receiver.

        TESLA (and GNSS PRF ranging broadly) requires the receiver maintain coarse time synchronization with an onboard, GNSS-independent clock~\cite{andersonTimeSync}.
        Therefore, the receiver will reject late signals arriving after the PRF ranging code's secret has been publicly revealed (at which point the PRF ranging code can be trivially computed by anyone).
        But there is still a window for a spoofer to submit small-delayed signals (i.e., arriving before the PRF ranging code's secret has been publicly revealed).
        Those spoofers must form their forgery by guessing the PRF ranging code or observe, process, and deliver a PRF ranging code by utilizing listening and transmitting antenna.
        The engineering task of building a listening and transmitting antenna is non-trivial.

        There exists a breadth of GNSS spoofing adversarial models, but for this work there is a useful binary classification: the adversary either engages in a Security Code Estimation and Replay (SCER) or it does not~\cite{Psiaki2016}.
        In an SCER attack, the adversary attempts to observe the unpredictable chips before submitting a forgery to the victim receiver.
        When the receiver does not engage in an SCER attack, it is called a Non-SCER adversary.
        This binary classification poses a convenience for the mathematical derivations of this work.

        \subsubsection{Non-SCER}

            The Non-SCER adversary makes no attempt to observe the cryptographic PRF ranging codes before submitting a forgery to the victim receiver.
            Therefore, were the Non-SCER adversary to submit a spoof of the form of \cref{eq:auth-signal}, it must guess its own replica and submit to the receiver:
            \begin{align}
                S^{\neg\text{SCER}} &= \sqrt P R^{\neg\text{SCER}} + N \mperiod \label{eq:s-nscer}
            \end{align}

            Suppose the Non-SCER adversary elects to manipulate its transmission by adding an adversarial noise term or varying $P$ over the aggregation interval.
            When the ranging signal is fed through the authentication filter, which is a match filter, the output enacts what amounts to an averaging filter with a large number of samples $FTW$ (e.g., exceeding $10230 \times 100$ for Galileo's E6-C).
            Because the Non-SCER adversary cannot observe $R^\text{PRF}$ before forming its forgery, each sample is weighted by $-1$ or $1$ with equal probability independent of the adversary's choice of the noise and $P$.
            Therefore, the expected contribution of this manipulation to the radio statistic of \cref{fig:radio} is exactly zero and its variance is the adversary's transmitted energy scaled by the filter gain.
            Thus, the manipulation cannot introduce bias but can only add power, which the receiver observes as degraded C/N$_0$ which \cref{sec:radio-model} bounds by refusing to authenticate below $30$ dB-Hz.

            A zero-mean, variance-bounded manipulation is not by itself sufficient to preserve the Gaussian tail, so the closed-form derivation of \cref{sec:awgn-non-scer} assumes the Non-SCER adversary submits a spoof of the form of \cref{eq:s-nscer} with constant $P$ and Gaussian $N$.
            Without any way to predict $R^\text{PRF}$, $R^{\neg\text{SCER}}$ will be a random guess of length $n$ resampled into $FT$.
            In this work, I assume the sampling $FT$ uniformly spans $n$ and that the small sampling quantity differences among chips poses no exploitable advantage to the Non-SCER adversary.
            Moreover, the same argument can be applied to the effect of measurement discretization and the associated gain control before the tracking loop.

            Utilizing the Non-SCER adversarial model, I show how to bound PMD and PFAs for PRF ranging authentication schemes.

        \subsubsection{SCER}

            The SCER adversary uses receiving antenna to estimate $R^\text{PRF}$ before submitting to its victim receiver.
            Because the adversary has access to measurements of $R^\text{PRF}$, the SCER adversary should have a better-than-50\% chance of guessing each chip, so cryptographic arguments will not be as helpful.
            As the SCER adversary utilizes better equipment, that probability will increase until it can perfectly estimate the individual chips and completely break the ranging authentication security.
            Unfortunately, cryptography is not useful for protecting against delays and delays affect the receiver's deduced PNT, so cryptographic security will not be obtainable for the SCER case.
            However, the argument can be offloaded on to whether the receiver can reject late signals and whether authorities can detect SCER receiving antenna under their jurisdiction.
            
            There are two principal ways to limit the effectiveness of SCER adversaries.
            The first principal way exploits how an SCER adversary must submit its spoof to the victim after a delay.
            It requires time for the SCER to observe the $R^\text{PRF}$, process its measurements, and transport its forgery to the victim receiver.
            Even with a no-latency SCER model, there is a fundamental limitation via speed of light.
            Therefore, the receiver can reject SCER-spoofed signals by rejecting late signals.
            However, the receiver cannot use the timing solution provided by GNSS or any other broadcast-only signal~\cite{andersonTimeSync}.
            This means that the receiver must utilize a GNSS-Independent clock to reject late signals.
            Then the challenge is the accuracy of this GNSS-independent clock: a nanosecond of uncertainty can mean about 3 meters of position uncertainty, which is not helpful or considered for this work.

            Instead, the second principal way exploits the low-power nature of the ranging signal.
            As the power of the signal decreases, the adversary's chip estimation process becomes more difficult, for which the adversary can compensate by increasing the effectiveness of its radio equipment.
            However, the adversary cannot increase the effectiveness of its radio equipment indefinitely without detection.
            As the signal's power decreases and the required SCER equipment increases, it should be easier for authorities on the ground to observe and destroy SCER spoofers.

            The mathematical derivations of this work utilize a limited SCER adversary called the Hard-decision SCER (HDSCER) adversary.
            The HDSCER adversary utilizes \cref{sec:chip-est-theory} to make hard decisions on each measured chip, ignoring potentially useful soft information (e.g., the posterior probability of the chip measurement).
            Since the HDSCER's only advantage over the Non-SCER is a $p$-accurate hard-decision estimate of the PRF ranging code, it will also not vary the power.
            Limiting to the HDSCER adversary will yield useful mathematically concise predictions similar to those by the Non-SCER adversary.
            Sadly, I am unaware of how to concisely extend the derivations herein to an SCER adversary that utilizes soft information.
            However, this assumption is relaxed in \cref{sec:pscer-adv}, where I show a simple soft-information-using SCER from \cite{scerpower,anderson2024gnsscrypto}, called PSCER, that demonstrates a small advantage to the HDSCER that manipulates the individual chip powers.
            This small advantage can be accounted for by adjusting the results of the HDSCER.
            There could be other soft-information adversaries that could do better.
            However, we approach the universal trade between an approximate problem with an exact solution versus an exact problem with an approximate solution.

            Utilizing the mathematical derivations for the HDSCER case, and the results from the PSCER case, I show how to predict the needed adversarial radio equipment, which would be useful to communicate to authorities on the ground so they know what to search for and destroy.

\section{Probability of False Alarm}\label{sec:nscer-pfa}

    To derive the PFA in the authentic for the threshold statistic from \cref{fig:radio}, let the authentic signal be
    \begin{align}
        S^\text{auth} &= \sqrt P R^\text{PRF} + N \mperiod \tag{\ref{eq:auth-signal}}
    \end{align}
    Now I feed $S^\text{auth}$ into the LTI filter $R^\text{PRF}$:
    \begin{align}
        Y_\text{PRF} \mid \text{auth} &= k_\text{PRF} \cdot R^\text{PRF}_- * S^\text{auth} \mcomma \nonumber \\
        &=  k_\text{PRF} \cdot R^\text{PRF}_- * (\sqrt P R^\text{PRF} + N) \mcomma \nonumber \\
        &=  k_\text{PRF} \sqrt P \cdot R^\text{PRF}_- *  R^\text{PRF} +  k_\text{PRF} \cdot R^\text{PRF}_- * N \msemi \nonumber \\
        Y_\text{PRF} \mid \text{auth} &=  1 +  k_\text{PRF} \cdot R^\text{PRF}_- * N \mperiod
    \end{align}
    Finally, specifying the distribution is a noise propagation exercise:
    \begin{align}
        N_{FT} &\sim \mathcal{N}\left(0, \frac{1}{FT} \frac{\sigma^2}{P}\right) \msemi \\
        Y_\text{PRF} \mid \text{auth} &= 1 + N_{FT} \mperiod \label{eq:pfa-1}
    \end{align}

    Because of the low-power nature of GNSS signals, the receiver will need to aggregate many of these statistics.
    Since \cref{eq:pfa-1} presents a normal distribution, taking the average of $W$ simply changes the variance of the noise term:
    \begin{align}
        N_{FTW} &\sim \mathcal{N}\left(0, \frac1W \frac{1}{FT} \frac{\sigma^2}{P}\right) \msemi \\
        Y_\text{PRF,W} \mid \text{auth} &= 1 + N_{FTW} \mperiod \label{eq:pfa}
    \end{align}

    The expectation $Y_\text{PRF,W} \mid \text{auth}$ is $1$.
    Later, in \cref{sec:sec-nscer}, I derive that the expectation in the Non-SCER spoofing case as 0.
    Without any reason to favor the PMD or PFA, I select a midpoint threshold of 0.5 for $Y_\text{PRF,W}$.
    To compute the PFA,
    \begin{align}
        \text{PFA} &= \prob \left(Y_\text{PRF,W} < 0.5 \mid \text{auth}\right) \nonumber \mcomma \\
        &= \prob \left(1 + N_{FTW} < 0.5 \right) \nonumber \msemi \\
        \text{PFA} &= \cdf_{N_{FTW}}(-0.5) \mperiod
    \end{align}

    In the following \cref{sec:sec-nscer}, I compute the PMD based on the 0.5 threshold and provide a plot like \cref{fig:PMD-nscer} to aid with the selection of $W$.
    However, I do not provide such a plot in this section.
    Because of the expectation symmetry of the 0.5 threshold in the authentic and Non-SCER spoofing case, and because the variance increases in the spoofing case, the PFA will always be less than the PMD in this symmetric 0.5 threshold case.
    In the following \cref{sec:sec-nscer}, I enable one to select $W$ to meet a specific PMD and security level.
    So I note that, given the construction of this threshold, the PFA will always be a slightly less than the PMD from \cref{sec:sec-nscer}.

\section{Non-SCER PRF Ranging Security} \label{sec:sec-nscer}

    This section covers Non-SCER PRF ranging security by bounding the PMD.
    First, I derive the PMD for the Non-SCER case when making assumptions about the noise being AWGN.
    Then, I apply the derivations to the encrypted Galileo E6-C signal used by Galileo's SAS to determine the aggregation time required for a given PMD.

    \subsection{Non-SCER with AWGN Case} \label{sec:awgn-non-scer}

    To derive the PMD in the Non-SCER case, let the spoofed signal be
    \begin{align}
        S^{\neg\text{SCER}} &= \sqrt P R^{\neg\text{SCER}} + N \mperiod \tag{\ref{eq:s-nscer}}
    \end{align}

    Before passing $S^{\neg\text{SCER}}$ through the LTI filter $R^\text{PRF}$, it is helpful to understand the result of convolving $R^\text{PRF}_-$ and $R^{\neg\text{SCER}}$ before either is resampled.
    Suppose that the adversary guesses every single chip of $R^\text{PRF}$ incorrectly.
    Then the convolution of $R^\text{PRF}_-$ and $R^{\neg\text{SCER}}$ is $-n$ (without resampling).
    When the adversary guesses every chip of $R^\text{PRF}$, then they are perfectly correlated.
    The likelihood that the adversary guesses any chips follows a fair binomial distribution, see Appendix \cref{sec: binom desc} and \cref{thm:binom} for justification.
    With each chip guessed correctly, the correlation increases by 2.
    Therefore,
    \begin{align}
        B^{\neg\text{SCER}} &\sim \mathcal{B}(n, 0.5) \msemi \label{eq:B-0.5-n} \\
        R^\text{PRF}_- * R^{\neg\text{SCER}} &= \left(-n + 2 \cdot B^{\neg\text{SCER}} \right) \frac{FT}{n} \mperiod
    \end{align}

    Let
    \begin{align}
        g(b) &= \frac 1n (-n + 2b) \mperiod
    \end{align}
    Now, I feed $S^{\neg\text{SCER}}$ through the LTI filter $R^\text{PRF}$:
    \begin{align}
        Y_\text{PRF} \mid \neg\text{SCER} &= k_\text{PRF} \cdot R^\text{PRF}_- * S^{\neg\text{SCER}} \mcomma \nonumber \\
        =  k_\text{PRF} &\cdot R^\text{PRF}_- * (\sqrt P R^{\neg\text{SCER}} + N) \mcomma \nonumber \\
        =  k_\text{PRF} &\sqrt P \cdot R^\text{PRF}_- *  R^{\neg\text{SCER}} +  k_\text{PRF} \cdot R^\text{PRF}_- * N \msemi \nonumber \\
        Y_\text{PRF} \mid \neg\text{SCER} &= g(B^{\neg\text{SCER}}) + N_{FT} \mperiod  \label{eq:yscer-after-dist-simplify} 
    \end{align}
    $Y_\text{PRF} \mid \neg\text{SCER}$ is simply the sum of a linearly transformed binomial distribution and the noise-propagated normal distribution.

    The receiver will aggregate $W$ of these identically distributed statistics to make an authentication determination.
    This means that the CLT can aid designing system parameters by providing closed-form expressions of the distributions.
    After performing a search of system parameters utilizing the CLT, one can then verify the system parameters with the actual distributions.

    To compute the $\mathbb{E}\left[Y_\text{PRF} \mid \neg\text{SCER}\right]$, noting that $g$ is a linear function:
    \begin{align}
        \mathbb{E}\left[Y_\text{PRF} \mid \neg\text{SCER}\right] &=  \mathbb{E}\left[g(B^{\neg\text{SCER}}) + N_{FT}\right] \mcomma \nonumber \\
        &= \mathbb{E}\left[g(B^{\neg\text{SCER}})\right] \mcomma \nonumber \\
        &= g(\mathbb{E}\left[B^{\neg\text{SCER}}\right]) \mcomma \nonumber \\
        &= g\left(\frac{n}{2}\right) \mcomma \nonumber \\
        \mathbb{E}\left[Y_\text{PRF} \mid \neg\text{SCER}\right] &= 0 \mperiod \label{eq:e-nscer}
    \end{align}
    And to compute the variance:
    \begin{align}
        \mathbb{V}\left[Y_\text{PRF} \mid \neg\text{SCER}\right] &=  \mathbb{V}\left[g(B^{\neg\text{SCER}}) + N_{FT}\right] \nonumber \\
        &=  \mathbb{V}\left[g(B^{\neg\text{SCER}})\right] + \mathbb{V}\left[ N_{FT}\right] \nonumber \\
        &=  \mathbb{V}\left[g(B^{\neg\text{SCER}})\right] + \frac{1}{FT} \cdot \frac{\sigma^2}{P} \nonumber \\
        &= \frac{4}{n^2} \mathbb{V}\left[B^{\neg\text{SCER}}\right] + \frac{1}{FT} \cdot \frac{\sigma^2}{P} \nonumber \\
        &= \frac{4}{n^2} \frac{n}{4} + \frac{1}{FT} \cdot \frac{\sigma^2}{P} \nonumber \\
        \mathbb{V}\left[Y_\text{PRF} \mid \neg\text{SCER}\right] &= \frac{1}{n} + \frac{1}{FT} \cdot \frac{\sigma^2}{P} \label{eq:V-nscer} \mperiod
    \end{align}

    Recall from \cref{sec:nscer-pfa} that $\mathbb{E}\left[Y_\text{PRF} \mid \text{auth} \right]$ = 1.
    Without any reason to favor the PFA and the PMD, I set the authentic versus spoof threshold at $Y = 0.5$.
    Now the goal should be to find system parameters with an acceptable PMD.

    The expectation and variance of $Y_\text{PRF}$ will be useful in performing an efficient search over $W$ to find candidate with a potentially acceptable PMD by employing the CLT.
    But after this search, one should directly compute the PMD.
    To compute the exact PMD, let
    \begin{align}
        B^\text{$\neg$SCER,W} &= \sum_{i=1}^W B^\text{$\neg$SCER}_i \nonumber \\
        B^\text{$\neg$SCER,W} &\sim \mathcal{B}(nW, 0.5) \mperiod \label{eq:B-0.5-nW}
    \end{align}
    Then,
    \begin{align}
        &\text{PMD} \mid \neg\text{SCER} = \prob (Y_\text{PRF,W} \geq 0.5 \mid \neg\text{SCER})\nonumber \mcomma \\
        &= \prob\left( \frac1W \sum_W \left(g(B^{\neg\text{SCER}}_i) + N_{FT} \right) \geq 0.5 \right) \nonumber \mcomma \\
        &=\prob\left( \frac1W \sum_W g(B^{\neg\text{SCER}}_i) + \frac1W \sum_W N_{FT} \geq 0.5 \right) \nonumber \mcomma \\
        &=\prob\left( g\left(\frac1W  B^{\neg\text{SCER,W}}\right) + N_{FTW} \geq 0.5 \right) \nonumber \mcomma \nonumber \\
        &= \sum_b \prob\left( g\left(\frac bW\right) + N_{FTW} \geq 0.5 \right) \cdot \prob\left( b = B^{\neg\text{SCER,W}} \right) \nonumber \\
        &= \sum_b \ccdf_{N_{FTW}}\left( 0.5 - g\left(\frac bW \right) \right) \cdot \prob\left( b = B^{\neg\text{SCER,W}} \right) \label{eq:PMD-nscer}
    \end{align}

    \Cref{eq:PMD-nscer} provides the exact PMD up to the computational precision of the normal Cumulative Density Function (CDF) and the effects of a any known adversarial advantage on the underlying PRF.
    I therefore present \cref{eq:PMD-nscer} as a design tool under the stated radio model.

    Several points inform my recommendation of a strong-margin PMD (e.g., 32-bit or 128-bit).
    First, the PMD is dominated by the Gaussian noise term rather than the binomial term.
    Second, PMDs matching bounds similar to those for security levels (e.g., a PMD of $2^{-128}$) extrapolate the AWGN model beyond the range any noise characterization can verify (a PMD of $2^{-128}$ corresponds to about 13 standard deviations of the radio statistic).
    Third, a strong-margin PMD isolates the spoofing hypothesis from the various other non-cryptographic effects on the signal that could throw a spoofing false alarm.
    Fourth, margins of at least 128-bit can account for the union-bound PMD degradation resulting from performing a spoofing check more than exactly once.
    However with a smaller margin resulting from a 32-bit PMD, the receiver should never attempt to correlate the same PRF ranging code segment more than exactly once (i.e., the receiver should never perform an acquisitions search with a PRF ranging code).
    Fifth, strong PMD margins will assist with the modeling weaknesses related to signal, radio, and PRF security models as these PRF ranging systems withstand future attacks.

    \subsection{Application to Galileo's E6-C} \label{sec:gal-e6-c-nscer}

\begin{figure}
            \centering
            \includegraphics[width=\linewidth]{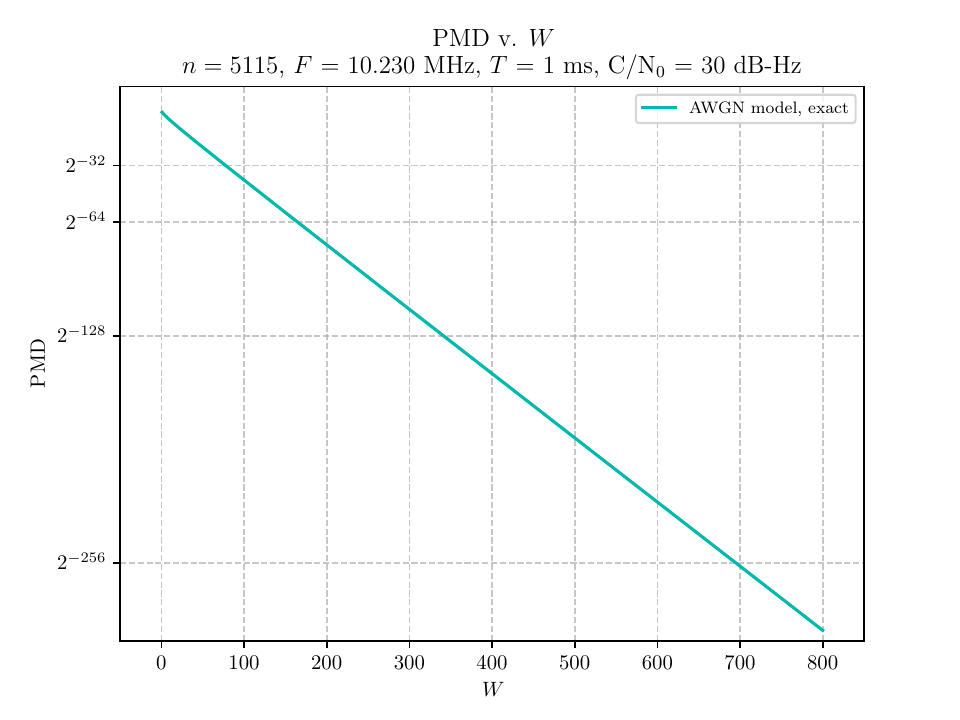}
            \caption{The PMD computed via \cref{eq:PMD-nscer}. \Cref{eq:PMD-nscer} yields $W=78$ and $W=342$ for PMDs of $2^{-32}$ and $2^{-128}$, respectively.}
            \label{fig:PMD-nscer}
        \end{figure}

        The Galileo E6-C signal will be used within Galileo's SAS to provide ranging authentication.
        The E6-C signal utilizes $n=5115$ ranging codes, and the appropriate Nyquist frequency is $F=10.230$ MHz.
        Once encryption is enabled, the ranging code will become a PRF under a commitment-based authentication protocol.
        Utilizing \cref{eq:PMD-nscer}, \cref{fig:PMD-nscer} provides the PMD varying $W$ integrally from $0$ to $800$.
        \cref{fig:PMD-nscer} was computed with SciPy utilizing the log survival function (log complementary CDF function) and the products within the sum of \cref{eq:PMD-nscer} utilizing the log product rule.
        Each of these steps was to minimize the losses from machine precision.

        Several applications utilize Galileo's E6-C PRF sequence to authenticate another band (e.g., E1-B).
        If the receiver does not track separately the E6-C PRF signal (e.g., via E6-B or via network), local environmental effects may require the receiver to conduct a search to find the correlation peak.
        For instance, this search could be required on the tracking states (e.g., carrier phase, doppler frequency, ranging code phase).
        Therefore, that receiver might try multiple times to find the PRF correlation.
        The net effect on the security model is that the adversary would be allowed multiple attempts to submit a potential forgery over a single range authentication determination.
        In that case, the minimum PMD should reflect a minimum security level of 128-bits, and so the receiver should aggregate 342 ms of data when making an authentication determination, based on \cref{fig:PMD-nscer}.

        If the E6-C authentication determination is aided with another signal that can provide accurate tracking states and the receiver is only allowed one attempt to make the authentication determination, then the security level can decrease provided the PMD is acceptable.
        This would be appropriate if another signal that is perfectly in phase is used to determine the E6-C tracking states, like the E6-B signal.
        Then 78 ms aggregations are sufficient, like in other similar applications where 32-bit security is acceptable~\cite{Andersonnavi.595}.

\begin{figure}
            \centering
            \includegraphics[width=\linewidth]{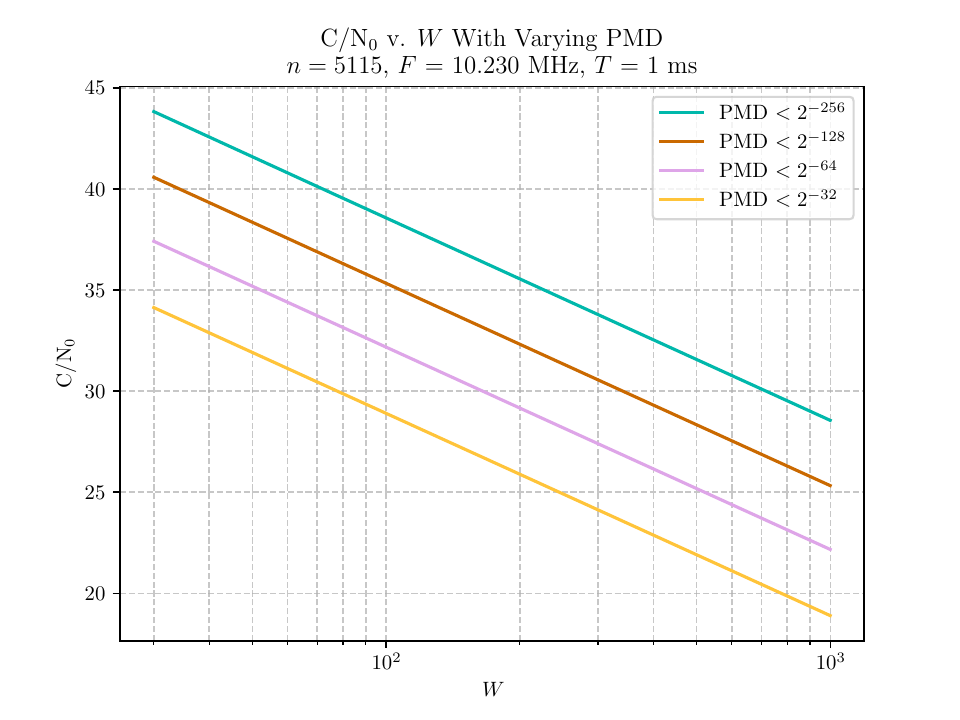}
            \caption{Minimum Noise Assumptions over Aggregation Time with varying security requirements utilizing the CLT approximations. Results below $W=30$ are not provided because this figure utilizes the CLT formulations.}
            \label{fig:cn0-v-W}
        \end{figure}

        Moreover, I expect that receiver guidelines will not recommend the minimum prescriptive boundaries from the previous paragraphs.
        Instead, I expect that the guidelines will round up to rounder numbers.
        For instance, from \cref{sec:gal-e6-c-nscer}, 78 ms could become 100 ms, and 342 ms could become 400 ms.
        If that is the case, then the initial noise assumption could be lowered further from 30 dB-Hz.
        \Cref{fig:cn0-v-W} demonstrates these trends utilizing the CLT approximations.
        Again, once a suitable candidate is found, \cref{eq:PMD-nscer} should be used to bound the PMD and assert authentication security.

\section{HDSCER PRF Ranging Security}\label{sec:hdscer-radio-gain}

    The Non-SCER adversary must guess the $R^\text{PRF}$, which explains the 50\% probabilities in \cref{eq:B-0.5-n,eq:B-0.5-nW}.
    But since the HDSCER adversary may measure $R^\text{PRF}$ before submitting to the victim receiver, the SCER adversary will have a better-than-50\% chance of guessing each chip.
    Because a better and better HDSCER receiving radio will eventually be able to estimate the PRF ranging code chips, this section offloads security arguments to authorities who can detect and search for this equipment.
    Ideally, the needed equipment will be large enough to be easily detected and destroyed.
    So in this section, I predict the receiving radio equipment required to perform an HDSCER spoof.

    With the HDSCER adversary, the adversary makes a hard decision on each individual chip.
    Therefore, the math for the HDSCER is identical to the Non-SCER, except that the binomial distribution probability $p$ is greater than 50\%:
    \begin{align}
        B^\text{HDSCER,W} &\sim \mathcal{B}(nW, p) \msemi \label{eq:B-p-nW} \\
        \text{PMD} \mid \text{HDSCER} &= \nonumber \\
        \sum_b \ccdf_{N_{FTW}}\left( 0.5 - g\left(\frac bW \right) \right) &\cdot \prob\left( b = B^{\text{HDSCER,W}} \right) \label{eq:PMD-hdscer}
    \end{align}

    The $nW$ chip agreements in \cref{eq:B-p-nW} are independent and identically distributed because the adversary's chip estimation errors are independent across samples at a constant SNR, which is an assumption about the adversary's receiver rather than the underlying PRF.
    So again, the CLT approximations can be helpful.
    In that case, \cref{eq:e-nscer,eq:V-nscer} can be extended:
    \begin{align}
        \mathbb{E}\left[Y_\text{PRF} \mid \text{HDSCER}\right] 
        &= g(\mathbb{E}\left[B^{\text{HDSCER}}\right]) \mcomma \nonumber \\
        &= g(np) \msemi \nonumber \\
        \mathbb{E}\left[Y_\text{PRF} \mid \text{HDSCER}\right] &= 2p-1 \msemi \label{eq:e-hdscer} \\
        \mathbb{V}\left[Y_\text{PRF} \mid \text{HDSCER}\right] &= \frac{4}{n}p(1-p) + \frac{1}{FT} \cdot \frac{\sigma^2}{P} \label{eq:V-hdscer} \mperiod
    \end{align}

    \subsection{Application to Galileo's E6-C}

        This section provides an example application of applying HDSCER PRF ranging security to Galileo's E6-C signal.
        Bounding the HDSCER adversary requires assumptions that are ultimately subjective.
        A better adversarial receiving radio will eventually break the security afforded by any ranging service.
        So this application, I assume that the HDSCER adversary is on the ground receiving the maximum specified power from the Galileo ICD and that the ground temperature is 300 K. 
        From the Galileo E6B/C Code Technical Note, the total received maximum power on the ground for E6-C is $-153$ dBW, after accounting for the 50\% power split to E6-C~\cite{GalicdE6}.
        Therefore, referenced to the chip rate $n/T = 5.115$ MHz (rather than the receiver's Nyquist rate $F=2n/T$), the precorrelation SNR is about $-16.26$ dB for the chip-estimating HDSCER adversary.

\begin{figure}
            \centering
            \includegraphics[width=\linewidth]{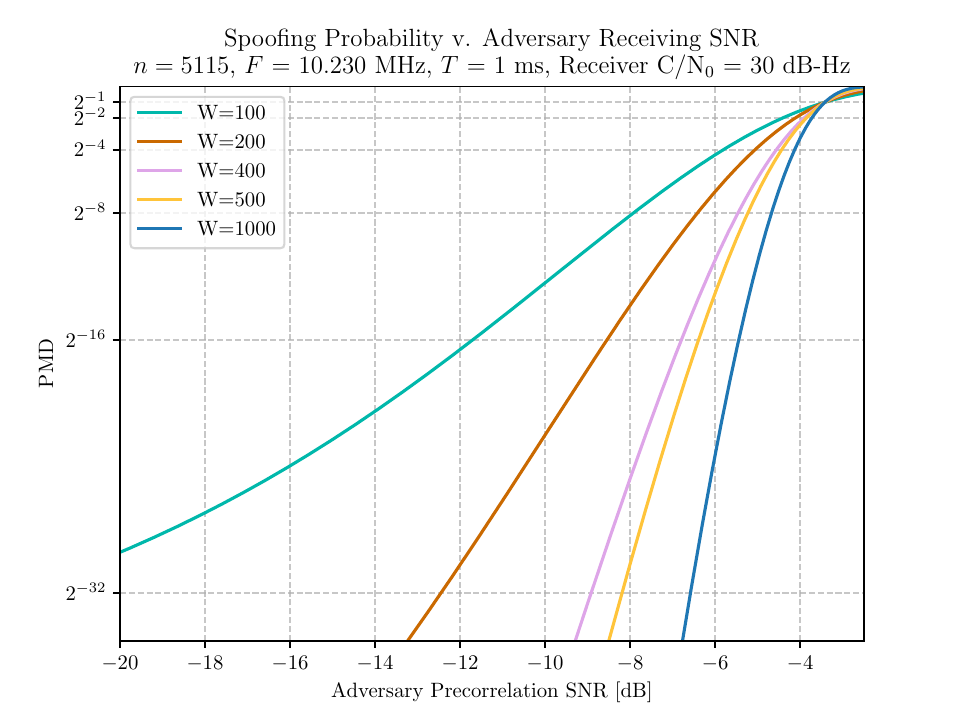}
            \caption{Comparison of HDSCER PMD with multiple $W$ against the adversarial precorrelation chip SNR.}
            \label{fig:hdscer-pmd-snr}
        \end{figure}

        \Cref{fig:hdscer-pmd-snr} provides the computed PMD given an adversarial chip-estimating SNR, generated from the CLT formulations from \cref{eq:e-hdscer,eq:V-hdscer} and assuming a chip estimation probability from \cref{eq:chip-est-correct} with a receiver C/N$_0$ of $30$ dB-Hz.
        I used the CLT formulations here because of the computational speed and the expected small deviation from \cref{eq:PMD-hdscer}.
        An assumed adversarial chip SNR of $-16$ dB is already breaking 32-bit security for $W=100$.
        However, given the PMD, it is very probable (i.e., exponentially likely) that the attack would be detected by the receiver after multiple $Y_\text{PRF,W}$ checks.

        An interesting behavior observable from \cref{fig:hdscer-pmd-snr} is that once the HDSCER adversary is more than 50\% likely to submit a forgery, increasing $W$ supports the adversary (hence the vertical order of the $W$ curves reverse in \cref{fig:hdscer-pmd-snr} on the right side).
        The effect of increasing $W$ removes uncertainty from the noise.
        At this inflection point, the adversary has a chip-estimation probability of 75\%, which from \cref{eq:e-hdscer} predicts that the expectation crosses the $Y_\text{PRF}=0.5$ decision boundary.
        Therefore, at that point, the shrinking noise effect from increasing $W$ makes the adversarial forgery probability better.
        This occurs when the chip SNR is $-3.42$ dB after solving \cref{eq:chip-est-correct} with 75\% (or when the adversarial gain is about $12.84$ dB considering the 300 K assumption).

        At this inflection point, the effects of increasing adversarial jamming no longer benefit the adversary.
        This means our conservative radio model from \cref{sec:radio-model} is no longer conservative.
        Increasing either $F$ or the assumed C/N$_0$ each benefit the adversary (rather than the reverse before).
        Therefore a conservative radio model would involve a high $F$ and receiver C/N$_0$.
        The net effect of this model's reversal past this inflection point is the adversary will be able to certainly and effectively break the security of the ranging authentication under a conservative radio model.
        Therefore, once the adversary estimates the chips with sufficient probability that its expectation of the spoofed $Y_\text{PRF}$ is above the decision boundary, the HDSCER completely breaks authentication security.

        \begin{figure}
            \centering
            \includegraphics[width=\linewidth]{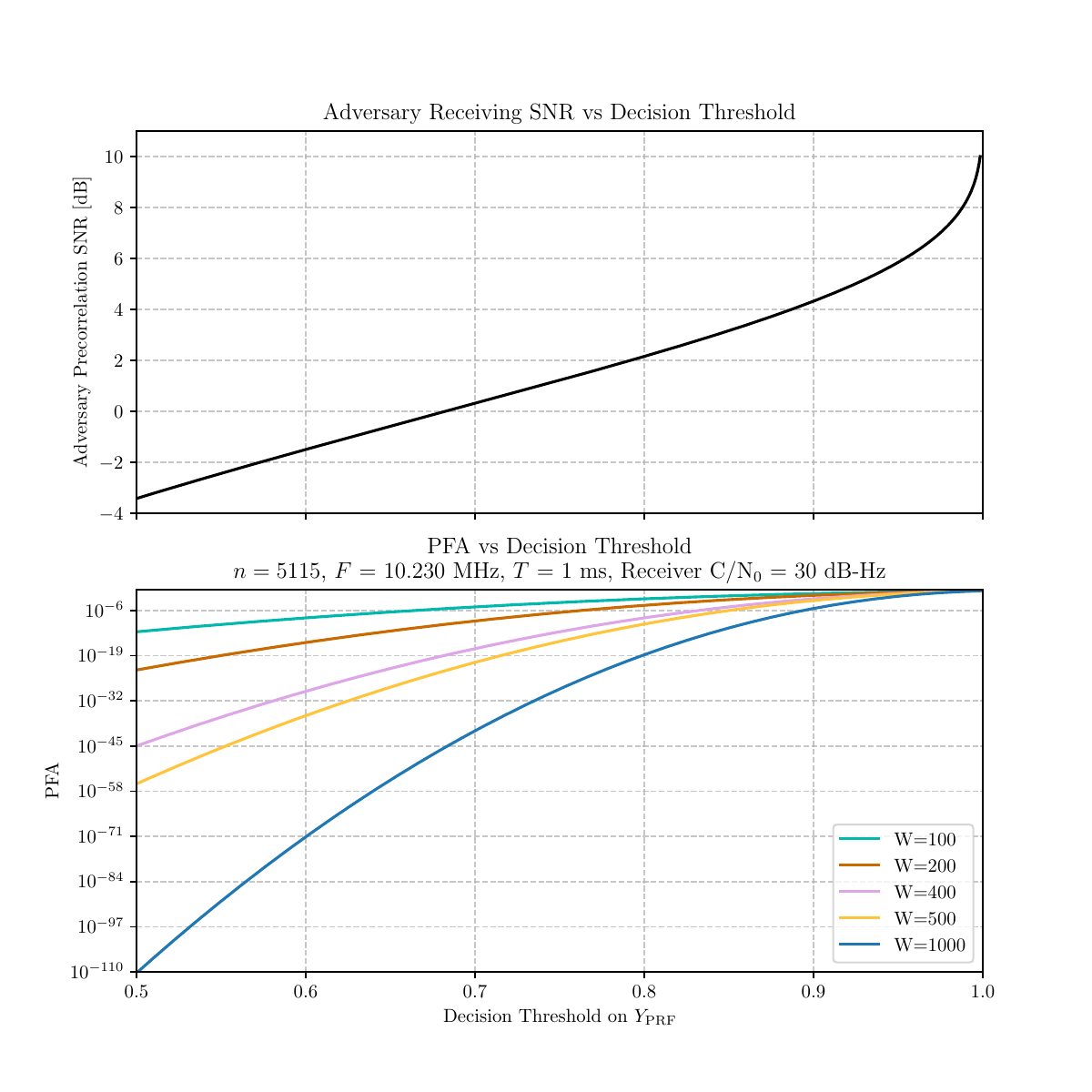}
            \caption{The effect of increasing decision boundary on $Y_\text{PRF}$ on the needed breaking Adversarial SNR and PFA.}
            \label{fig:adv-snr-pfa-y}
        \end{figure}

        To counteract this, one can increase the $Y_\text{PRF} = 0.5$.
        But that comes at the expense of the PFA.
        \Cref{fig:adv-snr-pfa-y} provides a plot of this trade off.
        After making a selection at the expense of PFA with \cref{fig:adv-snr-pfa-y}, \cref{fig:hdscer-pmd-snr} must be recalculated using the new decision boundary.
        Then, after computing an expected antenna gain to break the system, a description of that radio equipment of the systems can be communicated to authorities.
        Given the results from \cref{fig:hdscer-pmd-snr} already predict a radio of around 12 dB, this will be directional or phased array antenna that cannot be miniaturized, providing some level of assurance that such equipment could be detected and destroyed.

\section{Experimental Validation}

    To validate the PMDs derived in the previous sections, I perform Monte Carlo experiments.
    The previous sections, and their application to Galileo's E6-C, are constructed so that the PMD is unreasonably small to provide authentication security.
    These design goals pose the same challenge in simulation: an unreasonable number of trials is required to observe a single missed detection.
    So, in this section, the simulated C/N$_0$ is decreased substantially so that the mathematical models can actually predict simulated missed detections.

\begin{figure}
        \centering
        \includegraphics[width=\linewidth]{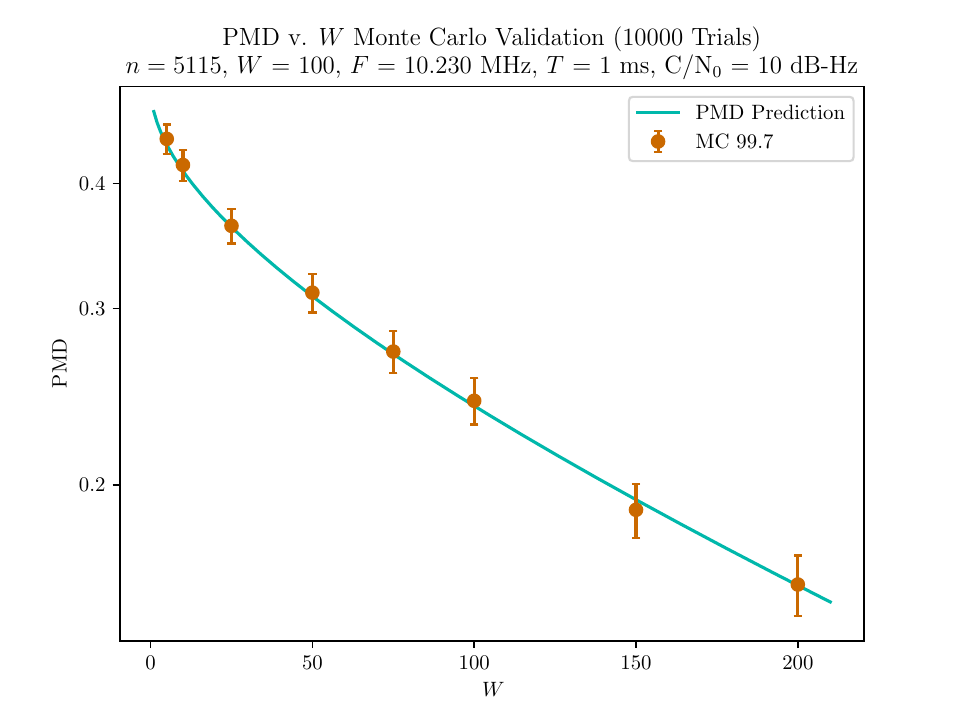}
        \caption{Monte Carlo results verifying \cref{eq:PMD-nscer}}
        \label{fig:validate-nscer}
    \end{figure}

    Based on how the experiments match the predictions, I validate \cref{eq:PMD-nscer} and its implementation, which also generates \cref{fig:PMD-nscer}.
    In the experiment, each trial involves a simulated baseband signal.
    The authentic and Non-SCER replicas are both chosen randomly and then resampled, with a simulated Non-SCER spoof fed through the simulated receiver filter and threshold-authentication determination.
    After conducting all the trials, the fraction of observed missed detections forms an estimator of the actual PMD.
    \Cref{fig:validate-nscer} compares the measured PMD and CLT 99.7\% confidence bounds against the predicted PMD.
    In our implementation, the same function is used to generate \cref{fig:PMD-nscer,fig:validate-nscer}.
    Based on how the experiments match the predictions, I validate \cref{eq:PMD-nscer} and \cref{fig:PMD-nscer}.

\begin{figure}
        \centering
        \includegraphics[width=\linewidth]{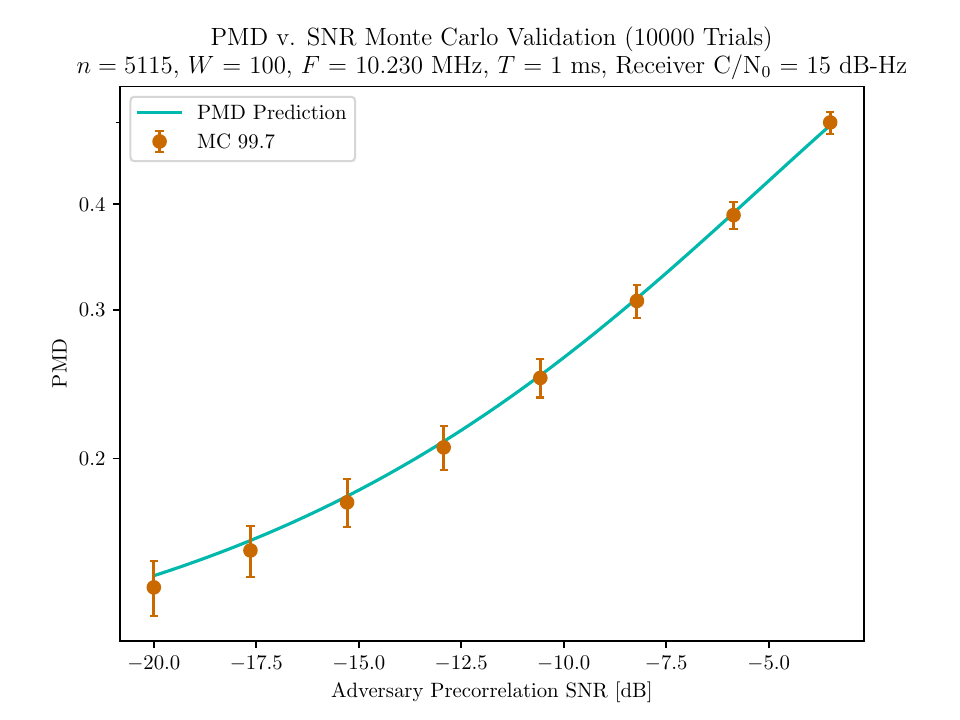}
        \caption{Monte Carlo results verifying \cref{eq:PMD-hdscer}}
        \label{fig:validate-hdscer}
    \end{figure}

    \Cref{fig:validate-hdscer} provides the results of a Monte Carlo experiment to verify \cref{eq:PMD-hdscer}.
    In the experiment, each trial involves a simulated baseband signal.
    The authentic replica is chosen randomly, and then I simulate the HDSCER's ability to estimate the authentic replica chips.
    Then both replicas are resampled with a simulated HDSCER spoof fed through the simulated receiver filter and threshold-authentication determination.
    After conducting all the trials, the fraction of observed missed detections forms an estimator of the predicted PMD.
    \Cref{fig:validate-hdscer} compares the measured PMD and CLT 99.7\% confidence bounds against the predicted PMD.
    In our implementation, the same function is used to generate \cref{fig:hdscer-pmd-snr,fig:validate-hdscer}.
    Based on how the experiments match the predictions, I validate \cref{eq:PMD-hdscer} and its implementation, which also generates \cref{fig:hdscer-pmd-snr}.

    \subsection{PSCER Soft Information Advantage}\label{sec:pscer-adv}

        While the HDSCER provides a mathematically concise pathway to predict the SCER security probabilities, this section provides simulated evidence of a slightly better SCER adversary.
        The HDSCER adversary ignores the soft information in its measurement and does not modify its power levels.
        Here, I define the PSCER adversary that sets the individual chip powers proportional to the measured posterior probability of each chip.
        But to counteract noticeable effects on the power, the PSCER adversary normalizes its output power over each PRF ranging code (hence ``proportional'').
        In effect, if the adversary is more certain on an estimated chip, it will provide that chip more power in its spoof.
        This adversary glosses over many effects, such as the effects of the radio's analog-to-digital gain control; however, the purpose of this section is simply to show the existence of a better adversary.

\begin{figure}
            \centering
            \includegraphics[width=\linewidth]{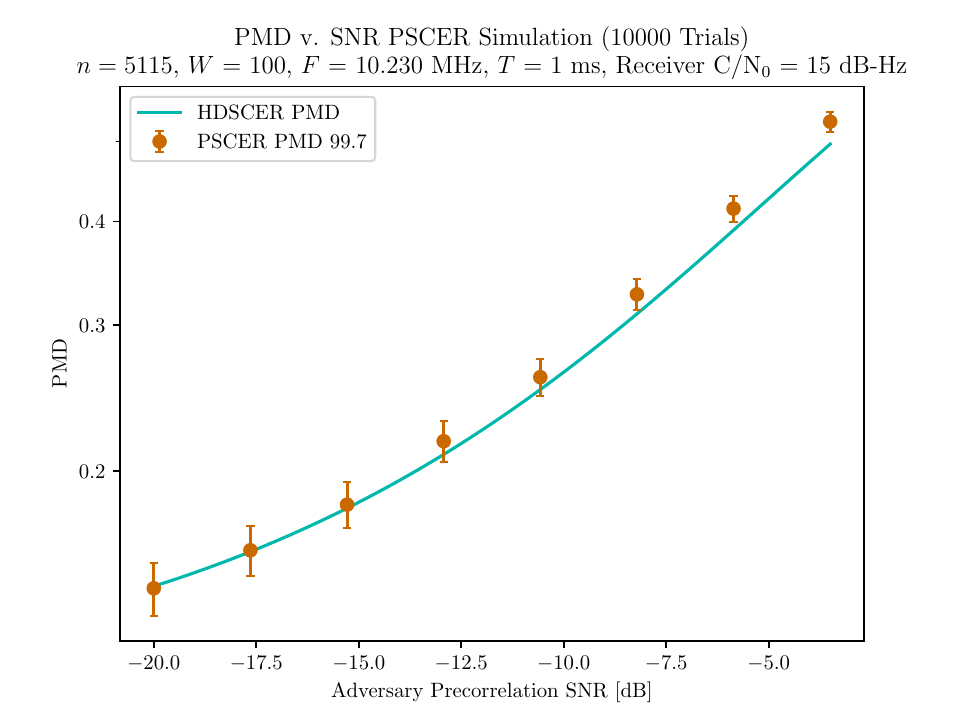}
            \caption{Simulated Results Demonstrating PSCER Advantage}
            \label{fig:experiment-pscer}
        \end{figure}

        \Cref{fig:experiment-pscer} provides a simulation showing a small advantage with the PSCER adversary over the HDSCER.
        Since the PSCER simulated PMD is above the HDSCER PMD, the PSCER has an advantage over the HDSCER.
        I provide these results to show that some performance improvement can be obtained by utilizing soft information, but not much.
        Looking at how far left the PSCER line is to the HDSCER line predicts the soft information advantage of approximately 0.6 dB.
        Therefore, one could account for this result by compensating the adversarial SNR of 0.6 dB within \cref{sec:hdscer-radio-gain}.
        However, this result still does not provide an upper bound on the potential SCER performance.

\section{Conclusion}

This work provides derived distributions of the authentication security of PRF GNSS ranging under several adversarial models.
With the derivations and predictions of this work, the GNSS designer can design a secure PRF ranging authentication service.
Under the Non-SCER model, one can utilize these derivations to specify how long a receiver must correlate PRF ranging codes to provide a specific PMD.
Under the HDSCER model, one can predict the needed spoofing radio equipment needed to break authentication security.
For Galileo's E6-C under a Non-SCER model, I conclude that a receiver must aggregate at most 100 or 400 ms of PRF Galileo E6-C to assert 32-bit and 128-bit security under Galileo's SAS, respectively.
Moreover, I predict that an SCER antenna with around a 12 dB gain antenna should be able to estimate the chips sufficiently to break authentication security of Galileo's E6-C PRF signal, neglecting the difficulty in quickly processing, transporting, and transmitting without detection either by a local authority or by the receiver's onboard GNSS-independent clock.

When a receiver utilizes a PRF pseudorange to authenticate another pseudorange, it will apply a pseudorange difference threshold to assert security of the other pseudorange.
When the PMD is bounded to standard cryptographic security levels (e.g., 128-bit), under a Non-SCER adversarial model, then the pseudorange difference threshold can be relaxed to increase the process's specificity (i.e., decrease PFA) and limit the false alarms resulting from environmental effects.
Under the Non-SCER model, even when a pseudorange is early or late, the adversary cannot produce a forgery up to the security level.
Therefore, the false alarms resulting from pseudorange delays (e.g., multi path, carrier frequency effects) can be removed by relaxing a pseudorange difference threshold without compromising sensitivity to spoofers.

\printbibliography

\appendices

\section{Non-SCER Adversary Chip Prediction Binomial Distribution Reduction to PRF Security} \label{app:proof}

In this appendix, I show that when the ranging code derives from a secure PRF with an unknown and uniformly random key, the chip agreement achieved by any efficient Non-SCER adversary is computationally indistinguishable from the binomial distribution, certifying \cref{eq:B-0.5-n}.
To do this, I first show that any Non-SCER adversary with an advantage in the ranging authentication attack game yields a PRF adversary with the same advantage in the PRF security attack game (\cref{thm: reduction}).
I then show that this implies the chip agreement count follows a binomial distribution up to a computationally indistinguishable difference.
The definition of PRF security and the following proof adapts Section 4.4 from \cite{boneh2023graduate}.
For clarity, $F$ and $S$ do not follow the notation from \cref{tab:notation} in this appendix.

A PRF $F$ is a deterministic algorithm with two inputs: a key $k$ from finite key space $\mathcal{K}$ and input $x$ from a finite input space $\mathcal{X}$.
Output $y = F(k, x)$ lies in finite output space $\mathcal{Y} = \{0, 1\}^n$.
Let $\textrm{Funs}[\mathcal{X}, \mathcal{Y}]$ denote the set of \emph{all} functions $f : \mathcal{X} \rightarrow \mathcal{Y}$.

\subsection{PRF Attack Game}\label{sec: attack game prf}
For a given PRF $F$, defined over $(\mathcal{K}, \mathcal{X}, \mathcal{Y})$, and for a given adversary $\mathcal{A}^\textrm{PRF}$, let there be two experiments: Experiment 0 and Experiment 1, indexed by $\beta = 0, 1$.
\begin{itemize}
    \item The challenger selects $f \in \textrm{Funs}[\mathcal{X}, \mathcal{Y}]$ as follows:
    \begin{itemize}
        \item if $\beta = 0$: $f \xleftarrow{\textrm{rand}} \textrm{Funs}[\mathcal{X}, \mathcal{Y}]$;
        \item if $\beta = 1$: $k \xleftarrow{\textrm{rand}} \mathcal{K}, f \leftarrow F(k, \cdot)$.
    \end{itemize}
    \item The adversary submits up to $Q$ queries $x_i\in \mathcal{X}$ to the challenger to assemble and utilize $y_i = f(x_i) \in \mathcal{Y}$.
    \item The adversary computes and outputs $\hat \beta \in \{0, 1\}$.
\end{itemize}
Let $W_\beta$ be the event that $\mathcal{A}^\textrm{PRF}$ outputs $1$ in Experiment $\beta$.
Then we define $\mathcal{A}^\textrm{PRF}$'s advantage with respect to F as 
\begin{align}
    \textrm{PrfAdv}[\mathcal{A}^\textrm{PRF}, F] = \left| \prob[W_0] - \prob[W_1] \right| \mperiod
\end{align}

A PRF $F$ is secure if for all efficient adversaries $\mathcal{A}$, the value of $\textrm{PrfAdv}[\mathcal{A}, F]$ is negligible.

\subsection{Non-SCER PRF Ranging Attack Game}\label{sec: attack game prf range}

For a given adversary $\mathcal{A}^\textrm{$\neg$SCER}$, again, let there be two experiments: Experiment 0 and Experiment 1, indexed by $\beta = 0, 1$.
Fix a subset $S \subseteq \{0, ..., n\}$.
Let $\Agree(R, R')$ with $R, R' \in \{-1, 1\}^n$ be the total number of chip agreements between $R$ and $R'$.
And let $g$ bitwise transform the binary string bit domain $\{0, 1\}$ into the signal chip domain $\{-1, 1\}$.
\begin{itemize}
    \item The challenger selects $f \in \textrm{Funs}[\mathcal{X}, \mathcal{Y}]$ as follows:
    \begin{itemize}
        \item if $\beta = 0$: $f \xleftarrow{\textrm{rand}} \textrm{Funs}[\mathcal{X}, \mathcal{Y}]$;
        \item if $\beta = 1$: $k \xleftarrow{\textrm{rand}} \mathcal{K}, f \leftarrow F(k, \cdot)$.
    \end{itemize}
    \item The adversary submits up to $Q$ queries $x_i\in \mathcal{X}$ to the challenger to assemble and utilize $y_i = f(x_i) \in \mathcal{Y}$
    \item The adversary commits to $R^\textrm{$\neg$SCER} \in \{-1, 1\}^n$.
    \item The adversary provides a fresh $x_f \not\in \{x_1, ..., x_Q\}$ to the challenger to receive a $R^f = g(f(x_f))$.
    \item The experiment outputs $\hat \beta = 1$ when 
    \begin{align} 
        \Agree(R^\textrm{$\neg$SCER}, R^f) \in S \mperiod \label{eq: agree}
    \end{align}
\end{itemize}
Let $W_\beta$ be the event that the experiment outputs $\hat\beta = 1$ in Experiment $\beta$.
Then we define $\mathcal{A}^\textrm{$\neg$SCER}$'s advantage with respect to F as 
\begin{align}
    \textrm{PrfRangeAdv}[\mathcal{A}^\textrm{$\neg$SCER}, F] = \left| \prob[W_0] - \prob[W_1] \right| \mperiod \label{eq: prf range adv}
\end{align}

A PRF range is secure if for all efficient adversaries $\mathcal{A}^\textrm{$\neg$SCER}$, the value of $\textrm{PrfRangeAdv}[\mathcal{A}^\textrm{$\neg$SCER}, F]$ is negligible.

\subsection{PRF Ranging Security Reduction}

\begin{theorem}\label{thm: reduction}
    If $F$ is a secure PRF, then the PRF range is secure. That is, for every efficient Non-SCER PRF Ranging adversary $\mathcal{A}^\textrm{$\neg$SCER}$ that plays the attack game from Appendix \cref{sec: attack game prf range} with respect to $F$, there is a PRF adversary $\mathcal{A}^\textrm{PRF}$ that plays the attack game from Appendix \cref{sec: attack game prf} with respect to $F$, where $\mathcal{A}^\textrm{PRF}$ is an elementary wrapper around $\mathcal{A}^\textrm{$\neg$SCER}$, such that
    \begin{align}
        \textrm{PrfRangeAdv}[\mathcal{A}^\textrm{$\neg$SCER}, F] = \textrm{PrfAdv}[\mathcal{A}^\textrm{PRF}, F] \mperiod \nonumber
    \end{align}
\end{theorem}

\begin{IEEEproof}
    Let $\mathcal{A}^\textrm{$\neg$SCER}$ be an efficient Non-SCER adversary that plays the attack game from Appendix \cref{sec: attack game prf range} with respect to $F$.
    I describe a corresponding PRF adversary $\mathcal{A}^\textrm{PRF}$ that makes at most $Q+1$ queries as follows:
    \begin{itemize}
       \item $\mathcal{A}^\textrm{PRF}$ plays the role of $\mathcal{A}^\textrm{$\neg$SCER}$'s challenger in the attack game from Appendix \cref{sec: attack game prf range}, forwarding each of $\mathcal{A}^\textrm{$\neg$SCER}$'s queries to its own challenger and passing back the answers.
       \item After receiving $R^\textrm{$\neg$SCER}$ from $\mathcal{A}^\textrm{$\neg$SCER}$, $\mathcal{A}^\textrm{PRF}$ queries its own challenger on $x_f$, computes $R^f = g(f(x_f))$ and $\Agree(R^\textrm{$\neg$SCER}, R^f)$, and outputs $1$ if and only if $\Agree(R^\textrm{$\neg$SCER}, R^f) \in S$.
    \end{itemize}

    With this construction, $\mathcal{A}^\textrm{PRF}$ answers every query of $\mathcal{A}^\textrm{$\neg$SCER}$ using the same function $f$ that its own challenger sampled, and outputs $1$ under the same condition as the attack game from Appendix \cref{sec: attack game prf range}.
    Hence, for each $\beta \in \{0, 1\}$, the probability that the attack game from Appendix \cref{sec: attack game prf range} outputs $1$ in Experiment $\beta$ is precisely equal to the probability that $\mathcal{A}^\textrm{PRF}$ outputs $1$ in Experiment $\beta$ of the attack game from Appendix \cref{sec: attack game prf}.
    Therefore,
    \begin{align}
        \textrm{PrfRangeAdv}[\mathcal{A}^\textrm{$\neg$SCER}, F] = \textrm{PrfAdv}[\mathcal{A}^\textrm{PRF}, F] \mperiod \nonumber
    \end{align}
    $\mathcal{A}^\textrm{PRF}$ runs $\mathcal{A}^\textrm{$\neg$SCER}$ plus $O(n)$ additional work to compute the chip agreement; thus, $\mathcal{A}^\textrm{PRF}$ is efficient.

    Since $F$ is a secure PRF and $\mathcal{A}^\textrm{PRF}$ is efficient, $\textrm{PrfAdv}[\mathcal{A}^\textrm{PRF}, F]$ is negligible, and therefore so is $\textrm{PrfRangeAdv}[\mathcal{A}^\textrm{$\neg$SCER}, F]$.

\end{IEEEproof}

\subsection{Binomial Distribution of Chip Agreement}\label{sec: binom desc}

\begin{theorem}\label{thm:binom}
    Let $B \sim \mathcal{B}(n, 0.5)$ and $F$ be a secure PRF.
    Fix $S \subseteq \{0, ..., n\}$.
    Then for every efficient Non-SCER adversary $\mathcal{A}^\textrm{$\neg$SCER}$, in Experiment 1 of the attack game from Appendix \cref{sec: attack game prf range},
    \begin{align}
        \left| \prob\left[\Agree(R^\textrm{$\neg$SCER}, R^f) \in S \right] - \prob\left[B \in S\right] \right| \nonumber
    \end{align}
    is negligible.
\end{theorem}

\begin{IEEEproof}
    Consider Experiment 0, where $f \xleftarrow{\textrm{rand}} \textrm{Funs}[\mathcal{X}, \mathcal{Y}]$.
    Because $x_f \notin \{x_1, \dots, x_Q\}$, the value $f(x_f)$ is uniform on $\{0,1\}^n$ and independent of everything $\mathcal{A}^\textrm{$\neg$SCER}$ has seen.
    Therefore, $R^\textrm{$\neg$SCER}$, which is committed before $x_f$ is queried, is also independent of $g(f(x_f))$.
    Each chip within $R^\textrm{$\neg$SCER}$ agrees with the corresponding chip of $R^f$ independently with probability $0.5$, regardless of $\mathcal{A}^\textrm{$\neg$SCER}$'s choice, so $\Agree(R^\textrm{$\neg$SCER}, R^f)$ is distributed as $B$.
    Thus $\prob[W_0] = \prob[B \in S]$.

    In Experiment 1, $\prob[W_1] = \prob[\Agree(R^\textrm{$\neg$SCER}, R^f) \in S]$ by definition.
    Substituting both into \cref{eq: prf range adv} yields
    \begin{align}
        &\textrm{PrfRangeAdv}[\mathcal{A}^\textrm{$\neg$SCER}, F] = \nonumber \\
        & \quad \left| \prob\left[B \in S \right] - \prob\left[\Agree(R^\textrm{$\neg$SCER}, R^f) \in S \right] \right| \mperiod \nonumber
    \end{align}

    By \cref{thm: reduction}, the left-hand side is negligible and hence so is the right-hand side.

    This formally establishes this reduction to PRF security for a single code ($W=1$, \cref{eq:B-0.5-n}).
    The $W$-code aggregate follows by the same arguments applied independently to each of the $W$.
\end{IEEEproof}

\end{document}